# Auger recombination in self-assembled quantum dots: Quenching and broadening of the charged exciton transition

*Annika Kurzmann[1*], Arne Ludwig[2], Andreas D.Wieck[2], Axel Lorke[1], and Martin Geller[1]*

1. Fakultät für Physik and CENIDE, Universität Duisburg-Essen, Lotharstraße 1, Duisburg 47048, Germany

2. Chair of Applied Solid State Physics, Ruhr-Universität Bochum, Universitätsstr. 150, 44780 Bochum, Germany

In quantum dots (QDs) the Auger recombination is a non-radiative process, where the electron-hole recombination energy is transferred to an additional carrier. It has been studied mostly in colloidal QDs, where the Auger recombination time is in the ps range and efficiently quenches the light emission. In self-assembled QDs, on the other hand, the influence of Auger recombination on the optical properties is in general neglected, assuming that it is masked by other processes such as spin and charge fluctuations. Here, we use time-resolved resonance fluorescence to analyze the Auger recombination and its influence on the optical properties of a single self-assembled QD. From excitation-power dependent measurements, we find a long Auger recombination time of about 500 ns and a quenching of the trion transition by about 80 percent. Furthermore, we observe a broadening of the trion transition linewidth by up to a factor of two. With a model based on rate equations, we are able to identify the interplay between tunneling and Auger rate as the underlying mechanism for the reduced intensity and the broadening of the linewidth. This demonstrates that self-assembled QDs can serve as an ideal model system to study how the charge recapture process, given by the band-structure surrounding the confined carriers, influences the Auger process. Our findings are not only relevant for improving the emission properties of colloidal QD-based emitters and dyes, which have recently entered the consumer market. They are also of interest for more visionary applications, such as quantum information technologies, based on self-assembled quantum dots.

The optical properties of solid state quantum systems, such as self-assembled[1] or colloidal quantum dots (QDs)[2], are strongly influenced, and sometimes limited, by electron and hole interactions. Charge carriers in the vicinity of a dot lead to spectral broadening[3-5]. Furthermore, electron and hole interaction inside the dot enables Auger recombination, a non-radiative process, where the electron-hole recombination energy is transferred to an additional charge carrier[6,7]. The Auger process has been studied extensively in colloidal QDs[8-10], where a fast Auger recombination time in the range below 1 ns[6,11] quenches the radiative recombination. This quenching limits the efficiency of optically active materials and devices based on colloidal QDs, such as light-emitting diodes[12,13] or single photon sources[14-16]. The roles of both extrinsic effects (e.g. coupling to external carrier systems) and intrinsic processes (such as Auger recombination) are topics still under intensive investigation[17-19].

The Auger recombination time in self-assembled QDs can be extrapolated to be in the nano- to microsecond range[20], orders of magnitude longer than for colloidal dots, so that this non-radiative process is of much lesser importance. Moreover, due to the highly ordered crystalline environment of self-assembled QDs, charging of nearby defects is strongly suppressed. This results in high quantum efficiency[21], almost transform-limited single photons[22-24], and a high photon indistinguishability[25,26].

Here, we use an electrically controllable self-assembled QD as a model system to study the competition between Auger recombination (intrinsic) and coupling to an external charge reservoir (extrinsic). By time-resolved resonance fluorescence (RF) spectroscopy, we determine an Auger recombination rate $\gamma_a = 2.3\ \mu s^{-1}$, which is of the same order of magnitude as the tunneling rate from the reservoir in our appropriately designed sample structure. This enables us to determine in detail how Auger recombination influences both the linewidth and intensity of the observed trion resonance. The experimental results are well accounted for by a rate model of the charge carrier dynamics. The model shows that a fast extrinsic process that replenishes the Auger ejected carrier can improve the optical properties, i.e. reduce the linewidth and increase the maximum intensity of the emitted light.

The investigated single InAs self-assembled QD is embedded in a (Al)GaAs heterostructure between a highly doped GaAs layer as a charge reservoir and a metallic top gate electrode (see methods and supporting information). Applying a suitable voltage to the gate allows us to controllably transfer a single charge from the reservoir to the QD via tunneling. Depending on the charging state of the dot, either the exciton transition or the trion transition is observed in RF.

In Fig. 1(a), the RF of the QD is shown as a function of gate voltage. Below a gate voltage of about $0.26\,V$, the Fermi-level in the back contact is below the lowest electron state in the QD and the QD is uncharged (sketched schematically in Fig. 1(a) top). At $V_g = 0.26\,V$, the first electron tunnels into the QD and RF of the trion transition $X^-$ is observed around $312.9\,THz$. The linear shift of the resonances is caused by the gate voltage-induce Stark effect[27].

The exciton RF intensity (red data points in Fig. 1(b)) exceeds the trion intensity (blue data points in Fig. 1(b)) by more than an order of magnitude for a broad range of laser powers. To test the assumption that the reduced intensity is caused by an Auger process (i.e. by the emission of the additional electron) we make use of the fact that the empty dot should exhibit resonance fluorescence of the uncharged exciton. Therefore, we apply a gate voltage, so that in equilibrium the dot is charged, and illuminate the sample with two lasers: The first laser (frequency 1) is in resonance with the trion energy (blue dot in Fig. 1(a)) and the second laser (frequency 2) is in resonance with the extrapolated (non-equilibrium) exciton energy (red dot in Fig. 1(a)). The spectrally resolved intensities of the exciton and the trion transition under two-laser excitation are shown in Fig. 1(c). We observe both the resonance fluorescence of the exciton and the trion at the same gate voltage $V_g = 0.318\ V$ (for more details, see supporting materials). At this gate voltage, the QD is charged with a single electron in equilibrium and no resonance fluorescence of the exciton transition is observed for single laser excitation (see Fig. 1(a)). Under two-laser illumination, the excitation of the trion together with the Auger recombination will result in an empty dot—see inset in Fig. 1(b). This non-equilibrium situation enables RF of the exciton transition until an electron tunnels back into the dot from the charge reservoir. The intensity of the exciton (red data points) exceeds the intensity of the trion (blue data points) by a factor of five in Fig. 1(c). The lower intensity of the trion compared to the exciton indicates that the Auger recombination rate $\gamma_a$ is higher than the tunneling rate $\gamma_{In}$, so that the dot is predominantly empty, i.e. in the non-equilibrium situation.

In the following, we will determine the Auger recombination rate in a time-resolved *m*-shot RF measurement with $m = 5 \cdot 10^5 - 10^7$ and a repetition rate of $20\ ms^{-1}$. The laser energy is adjusted so that a RF signal will occur at the trion transition. For each shot, we first prepare a singly-charged QD (see schematic in 2(b), left) by setting the gate voltage to $V_g = 0.318\ V$ and turning the laser illumination off for 25 μs. At $t = 0$, the laser is switched on and the QD can be excited in the trion state with the absorption rate $\gamma_{abs}$, see Fig. 2(b), middle. When the charged QD is in the excited state X⁻, Auger recombination can occur with a rate $\gamma_a$. The resulting electron emission switches the trion resonance off (see Fig. 2(b), right) until an electron tunnels from the reservoir back into the dot. The evolution from a QD that is charged with 100 % probability (at $t = 0$) to a steady-state situation, where the intensity is given by the competition between Auger emission and tunneling, is observed as an exponential decay with the relaxation rate $\gamma_m$. Figure 2(a) displays the corresponding signals for seven representative laser excitation powers, which determine the absorption rate $\gamma_{abs}$ and, hence, the probability $n$ for a trion inside the QD. The latter corresponds to the probability for the occupation of the upper level in a two-level system[28,29]. The conversion from laser power to $n$ is given by the red curve in Fig. 2(c) (see below). Note that in Fig. 2(a) the RF signal at $t = 0$ is normalized; the absolute intensity for $n = 0.003$ is much smaller than for $n = 0.5$. For $n = 0.003$ (low laser power), the time evolution of the normalized RF signal is nearly constant, because the dot is predominantly in the singly charged state (Fig. 2 (b), left). As a consequence, the Auger recombination is negligible. With increasing $n$, the decrease in the RF signal becomes more pronounced. For example, for $n = 0.1$, the normalized signal drops to a steady-state value of 0.5. At this laser excitation power, 50 percent of the measurements end in the situation where the electron was removed from the QD, due to the Auger recombination. For saturated excitation ($n = 0.5$) the RF is reduced by 80 % in the steady state. As sketched in Fig. 2(b), this signal quench depends on both the Auger recombination rate $\gamma_a$ and the tunneling rate $\gamma_{In}$. While the former is difficult to influence, the latter is tunable by the thickness of the tunneling barrier.

How the drop in the RF signal depends on these two parameters can be determined by a model based on rate equations. The time evolution of the normalized RF signal is given by the differential equation

$$\dot{P}_f(t) = \gamma_{In} P_{nf}(t) - n\gamma_a P_f(t), \tag{1}$$

where $P_{nf}$ and $P_f$ are the occupation probabilities for the empty dot (non-fluorescent state) and the charged state (fluorescent state), respectively.

The dependence of the upper state population in a two level system $n$ on the excitation power $p$ is given by the saturation curve[28, 29]

$$n(\Omega, \Delta\omega) = \frac{1}{2} \frac{\Omega^2 T_1/T_2}{\Delta\omega^2 + T_2^{-2} + \Omega^2 T_1/T_2}, \quad (2)$$

with the diagonal and off-diagonal damping constants $T_1$ and $T_2$, the Rabi frequency $\Omega \propto \sqrt{p}$ and the detuning $\Delta\omega$ ($\Delta\omega = 0$ in Fig. 2).

The initial condition $P_f(0) = 1$ is used to solve Eq. 1. We obtain

$$P_f(t) = \frac{\gamma_{In} + n\gamma_a e^{-\gamma_m t}}{\gamma_{In} + n\gamma_a}. \quad (3)$$

$P_f(t)$ directly reflects the measured transients in Fig. 2(a) with $\gamma_m$ given by

$$\gamma_m = \gamma_{In} + n\gamma_a. \quad (4)$$

To determine the Auger recombination rate, we use the tunneling rate $\gamma_{In} = 0.2 \, \mu s^{-1}$, derived from pulsed measurements of the excitonic RF (see supplementary and (30)). Using a fit of Eq. (3) to the data in Fig. 2(a) with the appropriate values of $n$ (red lines), an Auger recombination rate $\gamma_a = 2.3 \, \mu s^{-1}$ is obtained. This value is orders of magnitude smaller than Auger rates found in colloidal QDs. For example $\gamma_a = 0.1 \, ps^{-1}$ for CdSe dots of size $a = 2 \, nm$[11]. The discrepancy can be explained by a pronounced size dependence of the Auger rate, which was calculated to scale as $a^{-6.5}$ [20]. Using this scaling law and the lateral size of the present dots of $a \approx 10 \, nm$, we extrapolate an Auger rate of $2.8 \, \mu s^{-1}$ in good agreement with the value derived from Eq. (4). Note that the relaxation rate $\gamma_m$ is given by both Auger and tunneling rates. More importantly, a significant drop in the RF intensity can only be observed when the tunneling rate is similar to or smaller than the Auger recombination rate. Otherwise, an Auger ejected electron will be replaced instantaneously from the reservoir. Fig. 2(d) shows a calculation of the RF intensity in the steady state, $t \to \infty$, derived from Eq. (3):

$$P_f(\infty) = \frac{\gamma_{In}}{\gamma_{In} + n\gamma_a} = \frac{\gamma_{In}}{\gamma_m}. \tag{5}$$

Here, two limiting cases are displayed, corresponding to weak excitation $n = 0.01$ (red line) and saturation $n = 0.5$ (blue line). The influence of the tunneling rate on the trion RF intensity is clearly visible. For high tunneling rates $\gamma_{In} > 100\, \gamma_a$ the RF intensity is unaffected by Auger processes. This is the regime where most experiments have been performed on self-assembled QDs. In the opposite regime $\gamma_{In} < 0.001\, \gamma_a$, which applies to colloidal QDs, Auger processes have a strong influence on the optical properties and, for RF measurements, the intensity is completely quenched. In the present sample, a thick tunneling barrier was chosen, so that we are in the intermediate range $\gamma_{In} \approx \gamma_a$. Here, the RF intensity is affected by the laser power, which changes the normalized RF signal between 0.2 and $\approx 1$ (see vertical line in Fig. 2(d)).

In the following, we use a time-resolved two-color $m$-shot RF measurement to substantiate that the Auger processes indeed ejects an electron, leading to an empty QD, and that equilibrium will be reestablished by tunneling. As shown in Fig. 3, we first excite the QD in resonance with the trion transition until a steady state is obtained (see also Fig. 2(a)). Then the excitation frequency is changed to the exciton using a second laser. As seen in Fig. 3 (b), an RF signal of the exciton is observed after switching the laser frequencies at $t = 25\ \mu s$. This clearly shows that there is a non-vanishing probability to find the QD in an empty state after RF excitation of the trion.

For $t > 25\ \mu s$, an exponential decay in the exciton signal is observed with a relaxation rate of about $\gamma_r = 0.15\ \mu s^{-1}$. The good agreement of this value with the tunneling rate $\gamma_{In} = 0.2\ \mu s^{-1}$, obtained from pulsed gate voltage measurements, strongly supports our assumption that the dot is recharged by tunneling. The small discrepancy between $\gamma_r$ and $\gamma_{In}$ can be explained by optical blocking under RF excitation (see supporting information and (30)).

The interplay between electron tunneling and Auger recombination has a strong influence not only on the intensity of the trion resonance. It also affects its linewidth. Figure 4(a) shows the time dependent trion RF resonance from $t = 0$ (initialization, singly charged QD) to $t = 3 \, \mu s$ (charging state dependent on the interplay between Auger and tunneling processes). The data was recorded by measuring the time dependent RF for different detuning $\Delta\omega$ between the laser and the trion resonance, as shown in Fig. 4(b). Figure 4(c) displays two normalized spectra, taken at $t = 0$ and $t = 10 \, \mu s$, together with the corresponding Lorentzian fits. Clearly, the linewidth is increasing with increasing time as also summarized in Fig. 4(d). For very short times we observe a linewidth of 1.3 $\mu eV$, while for a steady state-situation ($t \rightarrow \infty$) the resonance is a factor of 1.6 broader.

This broadening can directly be explained by the influence of the detuning on the relaxation rate $\gamma_m$, which in turn depends on the trion population $n$, see Eq. (4). Equation (2) shows that $n$ decreases with increasing laser detuning $\Delta\omega$ and this leads to a reduced relaxation rate $\gamma_m$ and a reduced probability for Auger recombination. The measured relaxation rate, evaluated from fits to the transients in Fig 4(b), is summarized in Fig. 4(e) for different detunings $\Delta\omega$. The data can be described well by Eqs. (2) and (4), see blue solid line in Fig. 4(e). We observe a strongly reduced relaxation rate $\gamma_m$ at the edges of the transition compared to the center of the resonance.

The linewidth of the trion transition without an Auger process already depends on the laser excitation power, via the so-called power broadening. This linewidth follows from (31):

$$w(t=0) = \frac{2}{T_2}\sqrt{1+\Omega^2 T_1 T_2}, \quad (6)$$

where the Rabi-frequency $\Omega$ is given by the excitation power. We performed the measurement shown in Fig. 4(a) for different laser excitation power and determined the linewidth of the trion resonance at $t=0$ and $t=10$ µs. The measured linewidth of the transition at $t=0$ is shown as red data points in Fig. 4(f) together with a fit of Eq. 6 to the data (red line in Fig. 4(f)). The linewidth of the trion transition at $t=10$ µs can then be calculated without any adjustable parameter, using Eq. (2), (5) and (6) for $t \to \infty$:

$$w(t \to \infty) = \frac{1}{T_2}\sqrt{\frac{T_2^2 w(t=0)^2(\gamma_a + 2\gamma_{In}) - 4\gamma_a}{2\gamma_{In}}}. \quad (7)$$

It is plotted in Fig. 4(f) as blue line and is in good agreement with the measured data. The $T_2$ time can be determined from the linewidth at $t=0$ for small laser excitation power to be $T_2 = 975\ ps$, in agreement with previous estimates[28].

Our study allows the following conclusions: (i) Auger recombination leads not only to a reduced intensity but also an increased linewidth. (ii) To improve both, the required time to replenish the ejected carrier should be shorter than the Auger recombination time. This can easily be achieved in self-assembled QDs, e.g. by tunneling from a charge reservoir, however, it may be challenging for small colloidal dots having ps Auger recombination times. (iii) The Auger-induced broadening can be circumvented by measurements faster than the Auger recombination time.

In summary, we have studied the Auger recombination in a single self-assembled QD using time-resolved resonance fluorescence measurements. Our findings show that in samples that are only weakly coupled to a charge reservoir, the Auger process has a strong influence on the line shape of the trion transition: The linewidth increases and the maximum intensity decreases.

Using time-resolved RF and two color excitation we were able to determine both the tunneling rate and the Auger recombination rate, a quantity that was not well known for self-assembled dots. Furthermore, we have modeled the carrier dynamics in the QD using a simple rate equation which includes radiative recombination, Auger processes and tunneling. The calculations are in good agreement with our measurements and show that a fast tunneling time improves the emission characteristics of the QD. We believe that our findings are relevant for both self-assembled and colloidal QDs

**Methods**

The investigated sample was grown by molecular beam epitaxy (MBE) and resembles a field-effect-transistor structure[32,33], containing a layer of self-assembled InAs QDs. In detail, a 300 nm GaAs layer was deposited on a semi-insulating GaAs substrate, followed by a 50 nm heavily silicon-doped GaAs layer, which forms a nearly metallic back electrode and electron reservoir. A tunneling barrier, consisting of 15 nm GaAs, 10 nm $Al_{0.34}Ga_{0.66}As$ and 5 nm GaAs, was grown, which separates the InAs QDs from the doped back contact. The InAs QDs are formed by growing 1.6 mono-layers of InAs partially capped by 2.7 nm GaAs and flushed at 600 °C for 1 minute to shift the emission wavelength to ≈950 nm. They are further capped by a 27.5 nm GaAs layer, a 140 nm super lattice (35 periods of 3 nm AlAs and 1 nm GaAs) and 10 nm GaAs. The ohmic back contact is formed by AuGe and Ni evaporation and annealing (see supplemental material for more details of the sample). Transparent Schottky gates are prepared on the sample surface by standard optical lithography and deposition of 7 nm NiCr. On top of these gates, a zirconium solid immersion lens (SIL) is mounted to improve the collection efficiency of the QD emission[34]. A gate voltage applied between the top gate and the Ohmic back contact induces an external electric field and controls the charge state in the QD[35].

We use a confocal microscope setup in a bath-cryostat at a temperature of $4.2\ K$. For the RF measurements, the exciton (X) or trion (X⁻) transitions are driven resonantly by a linearly polarized and frequency stabilized tunable diode laser. In a confocal geometry, both laser excitation and QD emission are guided along the same path, using a 10:90 beam-splitter. Single QD resolution is archived by a 0.65 NA objective lens in front of the above mentioned SIL, giving a spot size of $1\ \mu m$. The emission of the QD is collected behind a polarizer, which is polarized orthogonally to the excitation laser and suppresses the laser light by a factor of $10^7$. The RF signal of the QD is detected by an avalanche photo diode (APD) and is recorded using a time-to-digital converter with a time resolution of $81\ ps$.

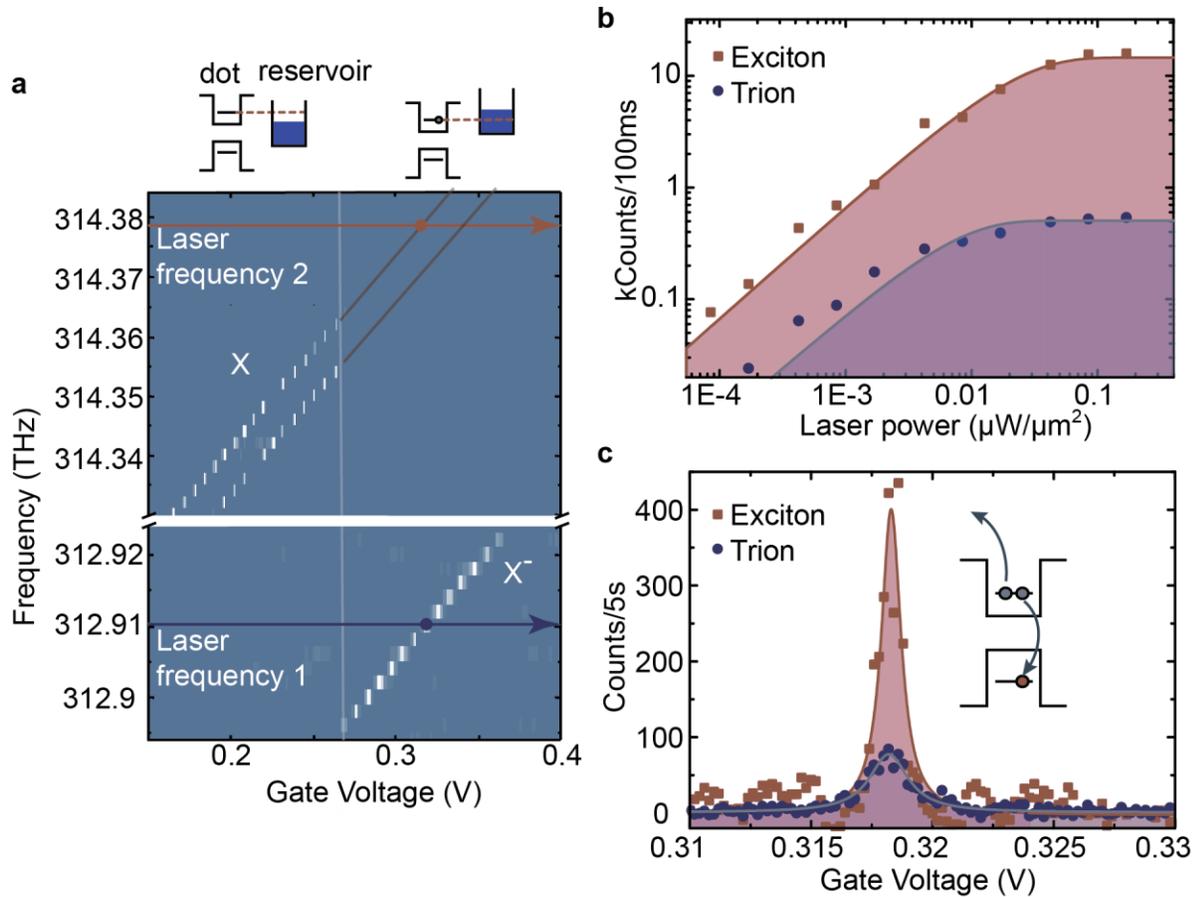

**Figure 1.** (a) Resonance fluorescence (RF) scan of the exciton (X) and trion (X⁻) for different laser excitation energies and gate voltages. The red and blue arrows indicate the voltage scan in (c), where we excite simultaneously on the trion (frequency 1) and exciton energy (frequency 2). The vertical white line marks the gate voltage where the Fermi energy in the charge reservoir is in resonance with the electron ground state in the dot, switching between the exciton and trion transistion. (b) Intensity of the RF signal of the trion (blue dots) and the exciton (red rectangles) for different laser excitation power. The exciton RF intensity exceeds the trion intensity by more than one order of magnitude. (c) Simultaneously measured RF spectra of the trion (blue rectangle) and the exciton (red dots). The observation of the exciton at this gate voltage is possible due to Auger recombination that emits the electron from the QD, leaving it in the uncharged ground state. This uncharged dot can now be excited at the exciton transition.

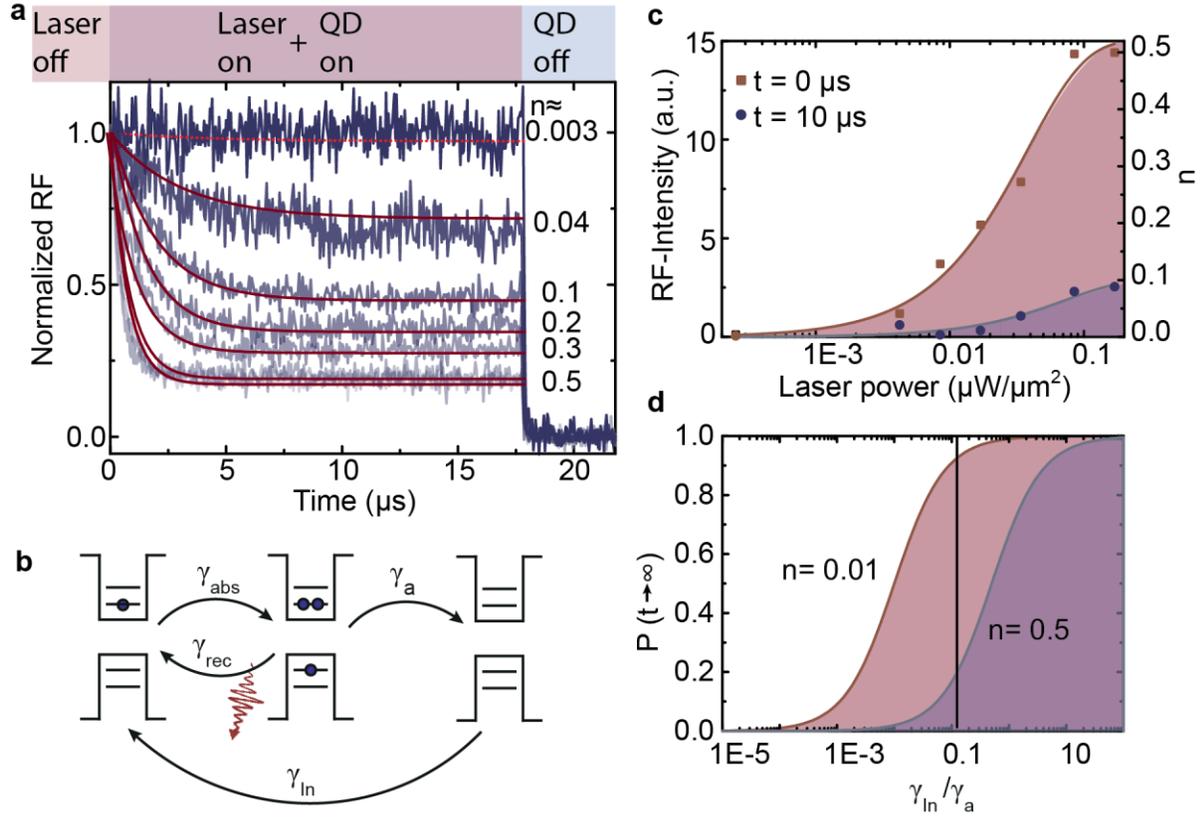

**Figure 2.** (a) Time-resolved RF of the trion transition. The laser is switched on at $t = 0$ and for increasing laser excitation power, increasing the average occupation of the dot with an charged trion $n$ (see Eq. (2) in the main text) is increased up to saturation $n = 0.5$. Increasing the average occupation $n$ results in a faster exponential decay and a smaller equilibrium amplitude, due to an increasing probability for an Auger recombination. (b) Schematic illustration of the different processes with rates for tunneling $\gamma_{In}$, Auger recombination $\gamma_a$, absorption $\gamma_{abs}$ and recombination $\gamma_{rec}$. (c) Intensity of the trion transition for different laser power at $t = 0$ (red rectangles) and at $t = 10\ \mu s$ (blue dots), taken without normalization from (a). (d) Calculations of the normalized steady state trion RF intensity for different ratios between tunneling and Auger rate $\gamma_{In}/\gamma_a$. For tunneling rates much smaller than the Auger rate $\gamma_{In}/\gamma_a \ll 1$ the RF signal is suppressed, while $\gamma_{In}/\gamma_a \gg 1$ yields a strong signal from the trion transition. The calculations are shown for high laser excitation power (blue line) and small laser excitation power (red line).

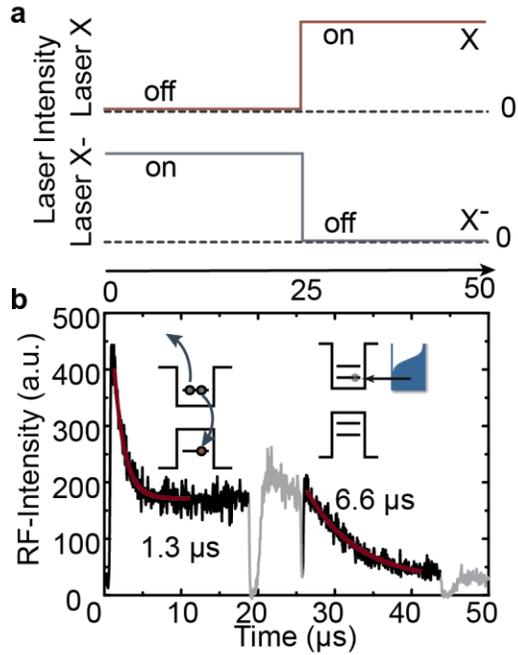

**Figure 3.** (a) Schematic picture of the laser pulse sequence for the two-color time-resolved measurement of the Auger recombination and tunneling rate. (b) The measurement shows an exponential decay of the RF trion signal for $t < 20$ μs, due to the Auger recombination for, after switching on the first laser labeled with X⁻. Switching on the second laser X to resonant excite the exciton for $t > 25$ μs yields a second exponential decay due to tunneling of an electron into the QD.

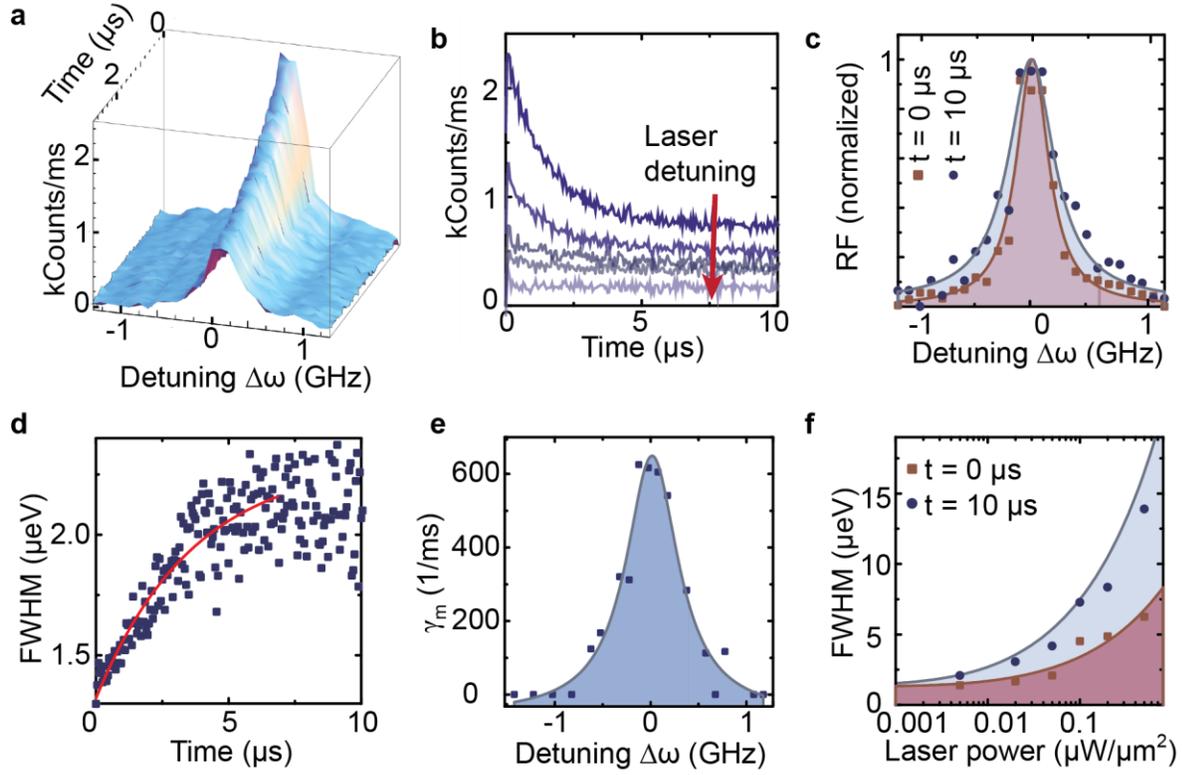

**Figure 4.** (a) Time-resolved measurement of the trion line after switching on the resonant excitation at $t = 0$ with a laser power of about $0.05 \frac{\mu W}{\mu m^2}$. For increasing time duration the line gets broadened and the maximum intensity is quenched by 60 percent. (b) The time-evolution of the line shapes in (a) has been determined by evaluation of transient for different laser detuning, showing different time constants for the relaxation rate $\gamma_m$. (c) Normalized trion resonance at $t = 0$ (red) and at $t = 10 \, \mu s$ (blue). (d) The linewidth of the trion transition increases from $1.3 \, \mu eV$ by a factor of 1.6 up to about $2.2 \, \mu eV$. (e) The relaxation rate $\gamma_m$ is plotted versus the laser detuning. We observe a decreasing relaxation rate for increasing detuning of the laser from the resonance maximum. (f) Linewidth of the trion transition at $t = 0$ (red) and at $t = 10 \, \mu s$ (blue) for different laser excitation power. The red line is a fit to the date and the blue line is calculated by the rate equation model (see main text).

## ASSOCIATED CONTENT

The supporting material includes measurement details, and measurements to the tunneling rates. This material is available free of charge via the Internet at http://pubs.acs.org.

## ACKNOWLEDGMENT

A.L. and A.D.W. acknowledge gratefully support of BMBF - Q.com-H 16KIS0109 and the DFH/UFA CDFA-05-06

## AUTHOR INFORMATION

**Corresponding Author**

* annika.kurzmann@uni-due.de

**Author Contributions**

The manuscript was written through contributions of all authors. All authors have given approval to the final version of the manuscript.